\documentclass[usenatbib]{mn2e}

\usepackage{amssymb}
\usepackage{amsmath}
\usepackage{epsfig}
\title{Comment on ``Biases in the Quasar Mass-Luminosity Plane''}
\author[Charles L. Steinhardt and Martin Elvis]
      {Charles L. Steinhardt and Martin Elvis\\
       IPMU, University of Tokyo, 5-1-5 Kashiwanoha, Kashiwa-shi, Chiba-ken, Japan 277-8583; \\ Harvard-Smithsonian Center for Astrophysics, 60 Garden St, Cambridge, MA 02138}

\date{\today}

\begin{document}

\let\la=\lesssim
\let\ga=\gtrsim
\def\case#1#2{\hbox{$\frac{#1}{#2}$}}
\def\slantfrac#1#2{\hbox{$\,^#1\!/_#2$}}
\def\onehalf{\slantfrac{1}{2}}
\def\onethird{\slantfrac{1}{3}}
\def\twothirds{\slantfrac{2}{3}}
\def\onequarter{\slantfrac{1}{4}}
\def\threequarters{\slantfrac{3}{4}}
\def\ubvr{\hbox{$U\!BV\!R$}}
\def\ub{\hbox{$U\!-\!B$}}
\def\bv{\hbox{$B\!-\!V$}}
\def\vr{\hbox{$V\!-\!R$}}
\def\ur{\hbox{$U\!-\!R$}}
\def\ion#1#2{#1$\;${\small\rm\@Roman{#2}}\relax}

\def\aj{\rm{AJ}}
\def\apj{\rm{ApJ}}
\def\apjl{\rm{ApJ}}
\def\apjs{\rm{ApJS}}
\def\mnras{\rm{MNRAS}}

\maketitle

\label{firstpage}




\medskip

In their paper ``Biases in the Quasar Mass-Luminosity Plane'',
\citet{Rafiee2010} (RH10) recalibrated Mg{\small II}-based virial mass
estimators for quasar black holes, and used these recalibrations to conclude
that the sub-Eddington boundary (SEB), described in \citet{Steinhardt2010a}
(SE10) is an artifact due to biases in the \citet{Shen2008} (S08) virial mass catalog.
The SEB is a bound restricting the most massive black holes over a wide range of
redshifts to radiate significantly below their Eddington luminosity. This is a
surprising feature of the quasar Mass-Luminosity plane, likely with major
implications, and so deserves to be tested rigorously.

Here we note, however, that the biases claimed by RH10: (1) are not
statistically significant; (2) imply non-virial motions in the quasar broad
line region so substantial that it is unlikely that the masses derived have
validity; (3) do not remove the SEB at most redshifts, but rather replace a bound restricting most quasars to radiate below Eddington at all redshifts with one restricting all quasars to radiate below Eddington at most redshifts.

\medskip
\noindent{\bf 1.  Statistical Significance:}

RH10 use an empirical relation between reverberation mapping masses and line
widths. They fit a slope, $\gamma$, to $M_{BH} \propto \textrm{FWHM}^{\gamma}$.
RH10 find a best-fit value of $\gamma = 1.27$ based on the MLINMIX\_ERR fitting
method of \citet{Kelly2007}.  The claim that the SEB disappears at $z \sim 2$
rests entirely upon this small value of $\gamma$.

RH10 argue that MLINMIX\_ERR is optimal. However the 18 different statistical
techniques they test produce values of $\gamma$ ranging from 1.21 to 3.95. The
value of $\gamma$ is clearly not robust but, rather, is highly sensitive to the
chosen fitting technique.  The error estimates provided by these different methods reflect 
this lack of robustness in that none gives a $\gtrsim 2\sigma$ deviation
below $\gamma$=2, the virial value\footnotemark, including the value
of $1.27 \pm 0.40$ preferred by RH10. We conclude that the current evidence, as
presented both by RH10 and in previous work, is consistent with the virial
approximation.

\footnotetext{Two methods give values $\sim$3$\sigma$ greater than 2. [OLS(X|Y)
and BCES(X|Y).]}

In the absence of an underlying physical model, corrections for substantial
non-virial motion are difficult to calibrate against the quite small ($N=29$)
reverberation mass sample. If RH10 are correct that $\gamma \neq 2$, many more
reverberation mass estimates will be required to produce compelling evidence,
robust with regard to fitting methodology, for a new choice of mass
estimation technique.

\medskip
\noindent{\bf 2. Non-Virial Motions:}

A key difference between this analysis and previous work is that RH10 do not
assume that motions in the quasar broad-line region are predominantly virial
(i.e. $M_{BH} \propto v_{virial}^2 \propto \textrm{FWHM}^2$).  While it is well known that broad emission lines have a
more triangular 'log' shape \citep{Petersonbook}, RH10 find that larger S08 BH masses are
systematically poorer approximations to gaussians, which could lead to a
mass-dependent bias in the S08 BH masses. 

Certainly, non-virial motions, mostly thought to be due to outflows, have long
been known in quasar spectra \citep{Gaskell1982,Wilkes1984}.  However, these are
strongest in the high ionization broad emission lines, notably in CIV, rather
than in low ionization lines such as MgII.  This ionization dependence is
expected for known sources of non-virial motion such as radiation pressure
\citep{Marconi2009}, as C{\small IV} comes from an inner region closer to the
ionizing continuum source and the central black hole, while Mg{\small II}, comes
from more distant material. In agreement with this expectation, the statistical
uncertainty in virial mass estimation is smallest for Mg{\small II}
\citep{Steinhardt2010c}.

If the Mass-FWHM relationship is as strongly non-virial ($\gamma$ = 1.27 rather
than $\gamma$ = 2) as RH10 claim, then this would call into question the
entire concept of virial mass estimation, which is predicated upon the
assumption that broad-line region motions are dominated by the gravitational
attraction of the central black hole. Without a physical model it is hard to
know how to correct for these motions, particularly given the strong dependence of the fit on the chosen statistical methodology.

Although there is no reason a priori that quasar luminosities must be able to reach Eddington, it is a boundary with physical meaning.  Using $\gamma = 2$, the most luminous quasars at every black hole mass reach but do not exceed Eddington at some redshift, and the highest-Eddington ratio quasars at every redshift reach but do not exceed Eddington.  Using $\gamma = 1.27$, as described below, for most choices of redshift no quasars reach Eddington.  That the virial approximation produces quasars reaching this boundary under a variety of conditions but never exceeding it implies that virial masses likely also have a physical meaning.  Non-virial masses do not share this property.

\medskip 
\noindent{\bf 3. RH10 Do See a Below-Eddington Boundary}: 

 The sub-Eddington boundary described in SE10 a boundary constraining quasars at most combinations of mass and redshift to lie below their Eddington luminosity, with the highest-mass quasars at each redshift being the most sub-Eddington and the lowest-mass quasars approaching $L_{Edd}$.  RH10 also see a below-Eddington boundary constraining quasars at most combinations of mass and redshift to lie below their Eddington luminosity, but with a different mass-dependence.

Even using the RH10 adjusted masses, their Figure 7 shows that the
highest-luminosity quasars at every mass at $z <$ 1.68 fall below Eddington,
with the shortfall monotonically increasing larger towards lower redshift. The tilt with
mass, however, is largely removed, although a few redshift bins
(e.g. 1.06$<$z$<$1.22) show a weak inversion of the SE10 result, with lower mass
quasars being more sub-Eddington than higher mass quasars.  The highest-redshift bin shown, $1.83 < z < 1.98$, again shows the lowest-mass quasars reaching Eddington but the highest-mass quasars falling short. 

The shortfall in RH10 Fig. 7 increases towards lower redshift down to the
0.76$<$z$<$0.912 bin, the lowest range plotted.  However, MgII is visible in SDSS
spectra down to z$\sim$0.4.  Fig. 1 shows the quasar $M-L$ plane
for 2 lower redshift bins, using both S08 (left) and RH10 (right) masses, down to this limit.  The additional shortfall using RH10 masses is more apparent in these bins, lying at least 0.5 dex below Eddington. A clear slope of increasing
deviation from Eddington at high mass is present in the lowest redshift bin (0.4$<$z$<$0.6).  In SE10, we showed the entire redshift range ($0.2 < z < 4.1$) allowed by our sample and techniques.  RH10 should do the same.

\begin{figure}
 \epsfxsize=3in\epsfbox{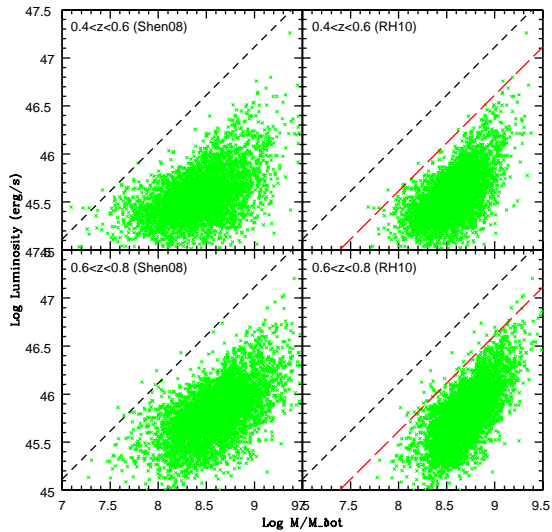}
\caption{The quasar mass-luminosity plane using S08 masses (left) and RH10
masses (right) at $0.4 < z < 0.6$ (top) and $0.6 < z < 0.8$ (bottom).  Using
S08 masses, the highest-mass quasars are all limited to lie below their
Eddington luminosities (black, short-dashed), while using RH10 all quasars are
sub-Eddington.  Using RH10 masses, quasars are increasingly sub-Eddington at
lower redshifts; quasars at $0.6 < z < 0.8$ approach closer to $\log (L/L_E) =
0.5$ (red, long-dashed) than at $0.4 < z < 0.6$.}
\label{fig:seb}
\end{figure}

The sub-Eddington boundary described in SE10 constrains quasars to lie below both a mass- and redshift-dependent luminosity limit.  Thus, adjusting the boundary to lie at Eddington also requires a combination of mass-dependent and redshift-dependent effects (SE10, \S~4).  RH10 give only a mass-dependent effect. 

\medskip
\noindent{\bf 4. Further Problems}

RH10 includes some misleading figures, captions, and text. Individually these
are not major, but the cumulative effect is unfortunate:
\begin{itemize}
\item In RH10 Fig. 1, the authors show quasars at all redshifts in their sample,
ranging from $0.76 < z < 1.98$, yet claim to draw the SEB described in SE10.  As
we demonstrate (SE10 Figs. 1, 3), the SEB is present in the quasar $M-L$ plane
only in narrow redshift intervals, and the location of the SEB moves upwards
with redshift. Hence the SEB is obscured by combining quasars spanning a wide
range of redshift (see e.g. SE10, Figure 1, and \citet{Kollmeier2005}).

\item In RH10 Fig. 7, Rafiee \& Hall note, as mentioned above, that their
SEB has disappeared ``at high redshift'', not in all panels.
Yet, in their abstract, they claim to have explained the entire effect as due to
biases in estimating masses.  Similarly, RH10 \S 4.1 claims that a change from
$\gamma = 2$ to $\gamma < 2$ has eliminated the sub-Eddington boundary; this is not what they have found.

\item In \S 4.1, RH10 quote SE10 as suggesting that any change in mass-scaling
relations should only shift the quasar locus in the $M-L$ plane.  In \S 4.5.2,
SE10 includes a full discussion of the topic, including allowing for the
possibility of a mass-dependent change in virial mass estimation.  SE10 also
considers previously proposed mass-dependent changes \citep{Onken2008,Risaliti2009}.

\end{itemize}

\medskip
\noindent{\bf 5. Discussion}

As discussed in a followup paper \citep{Steinhardt2010b}, the most puzzling
behavior in the mass-luminosity plane is that: (1) quasar accretion at a given
mass and redshift is apparently only possible within a narrow ($< 1$ dex);
luminosity range around a central, characteristic, sub-Eddington accretion rate and
(2) that characteristic accretion rate is time-dependent, decreasing
towards lower redshift.  

The key conclusion that quasar accretion rates are time-dependent yet
synchronized in host galaxies with quite different properties cannot be
explained merely by a mass-dependent bias in black hole mass estimation.
Instead a concidence between mass- and time-dependent biases is required.

The detailed quasar $M-L$ distribution, and our interpretation of that
distribution in an effort to build physical models, would be affected by solely
mass-dependent biases, so it is important to continue to improve our
understanding of virial mass estimation.  If motion in the broad-line region is
not predominantly virial, the technique may well need to be discarded rather
than patched.  However, Rafiee \& Hall do not claim statistically significant
deviations from virial masses, nor do their new calibrations remove either this
sub-Eddington behavior or other puzzling limits on quasar accretion.

The authors would like to thank John Silverman and Michael Strauss for valuable comments.

\bibliographystyle{apj}
\bibliography{ms}

\end{document}